# Broadband holography-assisted coherent imaging – towards attosecond imaging at the nanoscale

*Wilhelm Eschen [1,2,*], Sici Wang[1,2], Chang Liu[1,2], Michael Steinert[1], Sergiy Yulin[3], Heide Meissner[4], Michael Bussmann[4,5], Thomas Pertsch[1,3], Jens Limpert[1,2,3] & Jan Rothhardt [1,2,3]*

*Corresponding author, E-mail: wilhelm.eschen@uni-jena.de*

1. Institute of Applied Physics, Abbe Center of Photonics, Friedrich-Schiller-Universität Jena, Albert-Einstein-Str. 15, 07745 Jena, Germany
2. Helmholtz-Institute Jena, Fröbelstieg 3, 07743 Jena, Germany
3. Fraunhofer Institute for Applied Optics and Precision Engineering, Albert-Einstein-Str. 7, 07745 Jena, Germany
4. Helmholtz-Zentrum Dresden–Rossendorf, Bautzner Landstrasse 400, 01328 Dresden, Germany
5. Center for Advanced Systems Understanding (CASUS), Görlitz, Germany

**Abstract:**

In recent years nanoscale coherent imaging has emerged as an indispensable imaging modality allowing to surpass the resolution limit given by classical imaging optics. At the same time, attosecond science has experienced enormous progress and has revealed the ultrafast dynamics in atoms, molecules, and complex materials. Combining attosecond temporal resolution of pump-probe experiments with nanometer spatial resolution would allow studying ultrafast dynamics on the smallest spatio-temporal scales but has not been demonstrated yet. Unfortunately, the large bandwidth of attosecond pulses usually hinders high-resolution coherent imaging.

Here we present a robust holography-enhanced coherent imaging method, which allows combining high quality and high spatial resolution coherent imaging with a large spectral bandwidth. By implementing our method at a high harmonic source we demonstrate, for the first time, a spatial resolution of 34 nm (2.5 λ) in combination with a spectral bandwidth supporting a Fourier limited pulse duration of only 380 as. The method is single-shot capable, additionally retrieves the spectrum from the measured diffraction pattern, and is thus immune against shot-to-shot fluctuations. This paves the way for an ultrafast view on nanoscale dynamics e.g. ultrafast charge transfer or ultrafast spin currents being relevant for Petahertz electronics and future data storage.

**Introduction:**

Coherent diffractive imaging (CDI) techniques provide both amplitude- and phase-contrast and allow surpassing the resolution limits given by the manufacturing of the optics[1]. They rely on iterative phase retrieval algorithms to reconstruct the sample under investigation from its diffraction pattern. Especially in the XUV and X-ray range, where optics are lossy and challenging to fabricate, CDI has proven as a valuable methodology for high-resolution imaging[1–3]. In recent years CDI has seen great progress, the spatial resolution has been pushed to ~5 nm[4] and a variety of applications have been enabled ranging from 3D imaging of integrated circuits[5] or whole cells[6] to chemical sensitive imaging[4] of complex solid-state samples.

Lately, lensless imaging using table-top high harmonic sources has gained attention, since these sources allow performing experiments on an optical table that were so far only possible at large scale facilities[7,8,9,10]. At the same time, these sources can provide extremely broad spectra and pulses in the attosecond regime[11] with a natural synchronization to the ultrashort driving laser pulses making them suitable for ultrafast studies by pump-probe experiments[12]. As a result, attosecond physics has quickly evolved in recent years and allows us to answer fundamental questions in atomic[13], molecular[14], and solid-state physics[15,16]. Today, isolated attosecond pulses with µJ pulse energies are available[17] and attosecond pulses at free-electron lasers are emerging as well[18]. Combining the temporal resolution of attosecond science with nanoscale imaging would allow imaging of fastest dynamics at the nanoscale. However, lensless imaging in principle requires a monochromatic illumination and thus has so far been restricted to a narrow bandwidth and temporal resolutions in the few-ten femtosecond regime[19–21].

So far several sophisticated methods for the reconstruction of samples using broadband illumination have been demonstrated. Multi-shot techniques such as Ptychography[22,23] or the combination of Fourier transform spectroscopy with lensless imaging[24] have been demonstrated but appear difficult to be implemented in a pump-probe scenario, where additionally scanning of the pump-probe delay is required. Clearly, they are not capable of single-shot imaging. In contrast, most advanced phase retrieval algorithms[25] or deconvolution-based monochromatization of the measured diffraction pattern[26,27] demonstrated broadband coherent diffractive imaging from a single measured diffraction pattern with µm resolutions with visible light and hard X-rays. However, both approaches require precise knowledge about the spectrum of the illumination which might not always be accessible, in particular, if shot-to-shot fluctuations are present.

In this work, we present an alternative approach namely broadband holography-enhanced coherent diffraction imaging. In short, multiple reference structures around the sample provide additional information about the sample and the illumination, which is encoded in the measured diffraction pattern. This allows extracting the spectrum of the illuminating light source as well as a monochromatized and deblurred amplitude- and phase image of the sample. Subsequently, both can be used as input to a broadband iterative phase retrieval algorithm for further refinement of the image. The holographic input data significantly improves the reliability, convergence, speed, and image quality of the iterative phase retrieval.

In a proof-of-principle experiment, using a high harmonic source centered at 92 eV, a spatial resolution of 34 nm has been achieved with a bandwidth of 5.5 eV, which corresponds to a Fourier-limited pulse duration of only 380 as. The observed image features remarkable quality and wavelength-scale resolution (2.5 λ), which is close to the Abbe-limit of 25 nm, which demonstrates the capabilities of this approach.

The presented method is ideally suited for table-top sources and can be employed at synchrotrons and free-electron-lasers as well. It is single-shot capable, scanning free, and tolerates fluctuations of the spectrum, pointing, and power of the source. Thus, broadband holography-enhanced coherent imaging paves the way for many applications that desire an ultrafast view on dynamics and quantum effects at the nanoscale including ultrafast transport and transfer processes of energy[28], charges[15] or spins[29] which are the basis of next-generation electronics, data storage, energy conversion, and energy storage devices.

**Results**

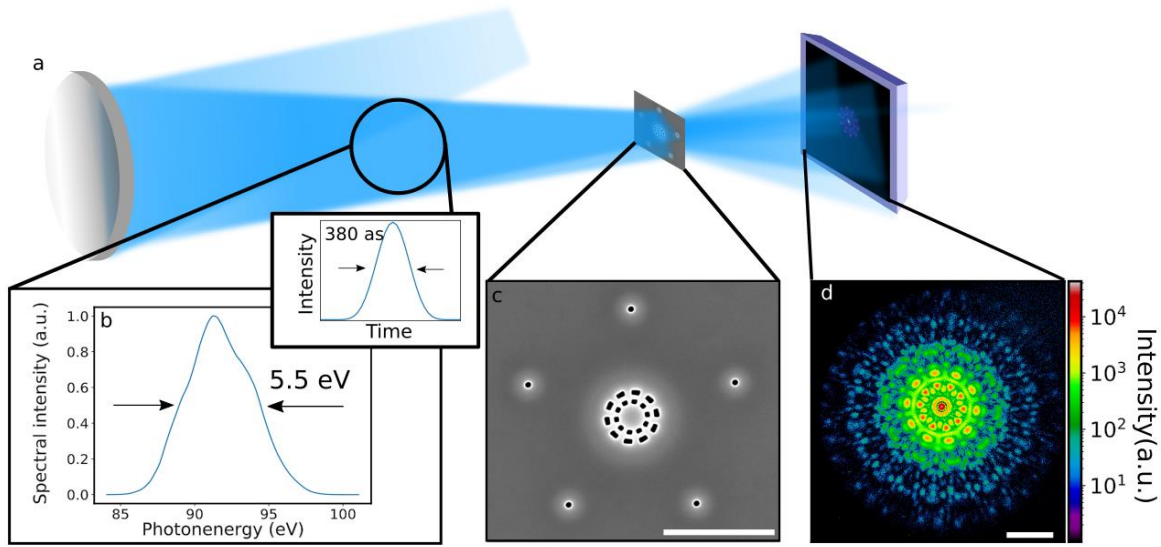

**Figure 1: Schematic representation of the experiment. a** A spatially coherent high-harmonic beam is focused on the sample using a broadband multilayer mirror. **b** The filtered spectrum possesses a bandwidth of 5.5 eV. The small inset shows the XUV-pulse in the time domain yielding a pulse duration of 380 as, assuming a flat spectral phase. **c** SEM image of the sample: A resolution-test pattern is located in the center and surrounded by five reference apertures (scale bar in c corresponds to 2 µm). **d** Measured (intensity) hologram of the sample in log-scale The scale bar in d corresponds to 10 µm$^{-1}$.

For the presented experiments we used a laser-driven, high-order harmonic (HHG) source which generates a broadband XUV spectrum up to 100 eV. From this very broadband spectrum, a bandwidth of 5.5 eV at a central photon energy of 92 eV is selected, which corresponds to a relative bandwidth λ/Δλ of 17 and a Fourier-limited pulse duration of 380 as (Fig. 1 **b**). A more detailed description of the XUV source and experimental geometry is given in the methods section. A Siemens star-like resolution test chart with a diameter of 1 µm, is placed in the focus of the XUV-beam and acts as a test sample (see Fig. 1 **c**). To provide holographic reference waves, five circular apertures with a diameter of 90 nm are placed evenly spaced around the sample on a circle with a radius of 2 µm, which is also known as Fourier transform holography[30] (FTH). The recorded hologram is shown in Fig. 1 **d**.

In order to extract a high-resolution image of the sample, we follow a two-step approach. First, we extract an amplitude and phase image of the sample from the hologram via FTH.

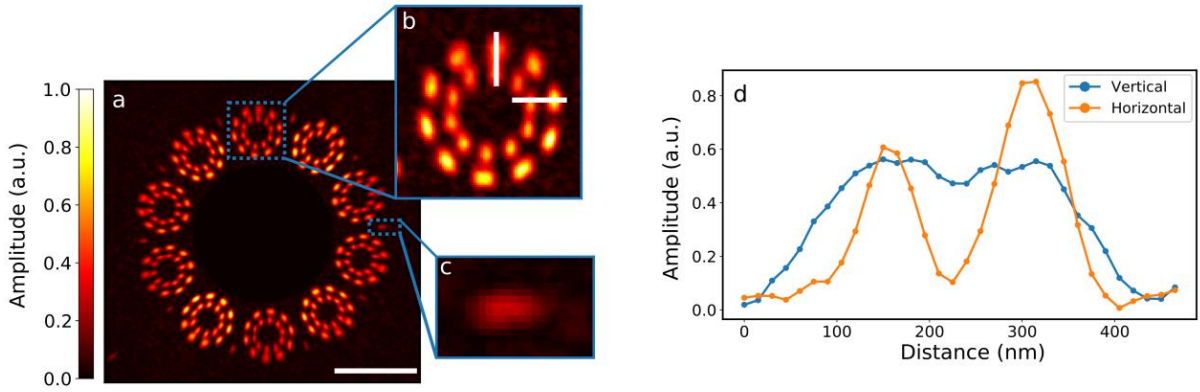

**Figure 2: Broadband Fourier transform holography. a** Fourier transform of the measured hologram shown in Fig. 1 **d**. The intense autocorrelation in the center is blocked and only the five cross-correlation terms and their corresponding complex-conjugates are shown. The scale bar represents 1.5 µm. Figure **b** shows a magnified version of the top cross-correlation. **c** shows a magnified version of a smeared cross-correlation of two reference apertures. **d** shows the vertical and horizontal lineouts indicated in **b**.

Thus a simple Fourier transform of the measured hologram (Fig. 1 **d**) yields a representation of the objects exit wave, further named object, as shown in Fig. 2 **a**. Since the object is surrounded by five reference pinholes, five reconstructed objects (cross-correlation of the sample with the reference) and their complex-conjugated are visible. A closer look at one of the reconstructions (Fig. 2 **b**) shows that the object is noticeably smeared, which can be attributed to the broad-bandwidth radiation used in the experiment.

A vertical and horizontal lineout along the features of the Siemens star is shown in Fig. 2 **d**. Clearly, the smearing effect (i.e. the spatial resolution) is anisotropic. The features along the horizontal direction are resolved, while the features along the vertical direction are blurred. This behavior is explained in more detail in the supplement. Mainly, the large distance between the reference hole and sample causes a large shift of the cross-correlation term with the wavelength. For a broadband source, this effect leads to continuous smearing along the pinhole-sample direction, whereas the smearing in perpendicular direction is much weaker.

In order to de-smear the image of the object, we utilize the diverse information provided by the five pinholes. In Fig. 2 **a** one can clearly see that each cross-correlation term is smeared in a different direction. Instead of averaging over all reconstructed objects[31], we thus combine the non-smeared spatial frequencies from all cross-correlation terms to a single de-smeared image. For doing so, we cut the five reconstructed objects (see Fig. 3 **a**) from the reconstruction and apply a Fourier transform on each of the objects (Fig. 3 **b**). In Fourier space, we apply a suitable filter on each object. Here, five binary, hourglass-shaped masks (small inset in Fig. 3 **b**) are used, which keep the high-resolution spatial frequency components along the axis perpendicular to the sample-pinhole direction and discard the low-resolution spatial frequency components.

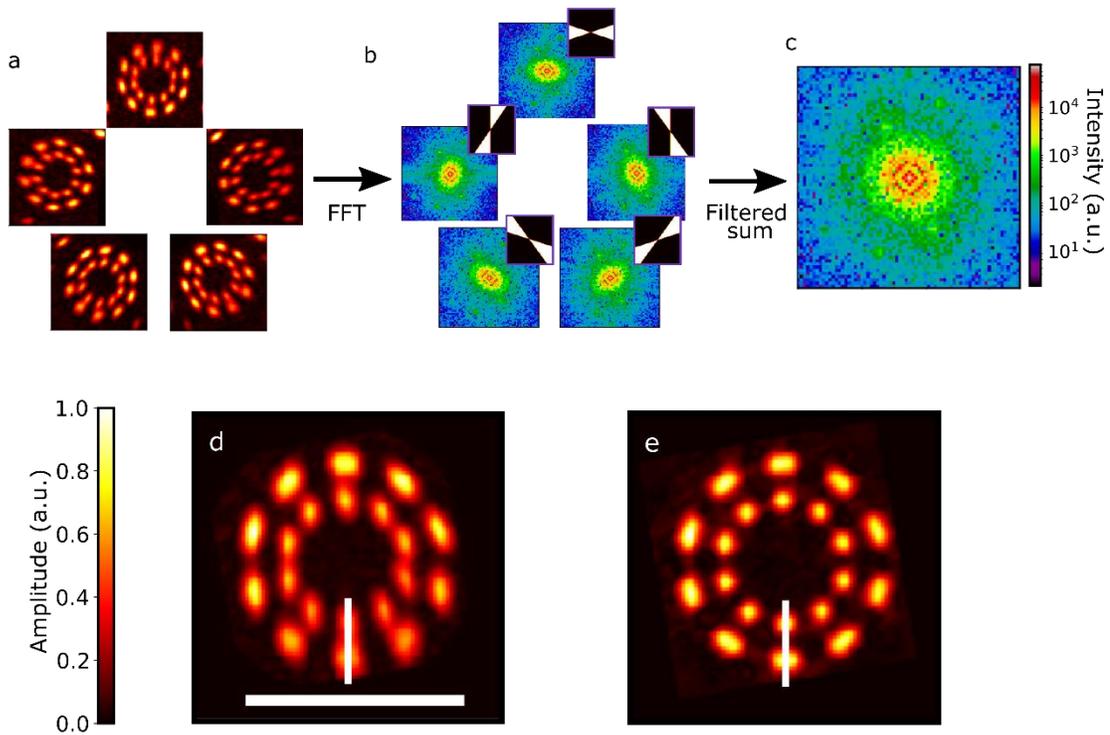

**Figure 3: Illustration of the broadband Fourier Transform Holography method. a** Each cross-correlation term is transformed into the Fourier space **b** where a binary filter is applied (small inset), which keeps the high-resolution spatial frequencies. The spatial frequencies are summed and yield a high-resolution Fourier space image which is shown in **c**. **d** shows an enlarged smeared cross-correlation term before the broadband FTH method was applied and **e** shows the result of the broadband FTH method. The scalebar in **d** corresponds to 1 µm.

The remaining high-resolution spatial frequency components are summed up (Fig. 3 **c**) and finally the improved object is obtained by an inverse Fourier transform. The resulting high-resolution reconstruction of the object (Fig. 3 **e**) shows a significantly improved image quality compared to the initial reconstruction (Fig. 3 **d**). A lineout for both reconstructions is shown in Fig. 5 and shows that two features separated by a distance of 90 nm that were so far not resolved, are now clearly distinguishable. A knife-edge test along the white line in Fig. 3 **d** suggests a resolution of 115 nm for standard FTH, which is in good accordance with the temporal coherence limit given by the bandwidth (110 nm). In contrast, the broadband-FTH method achieves a resolution of 65 nm along the white line in Fig. 3 **e**, which agrees well with the 60 nm limit imposed by the diameter of the reference aperture[32] (~70% of the diameter).

So far, we have demonstrated an improved resolution of 65 nm using broadband FTH, which beats the temporal coherence limit by a factor of 1.7. From theory this de-smearing factor can be as large as 5, more details can be found in the supplement.

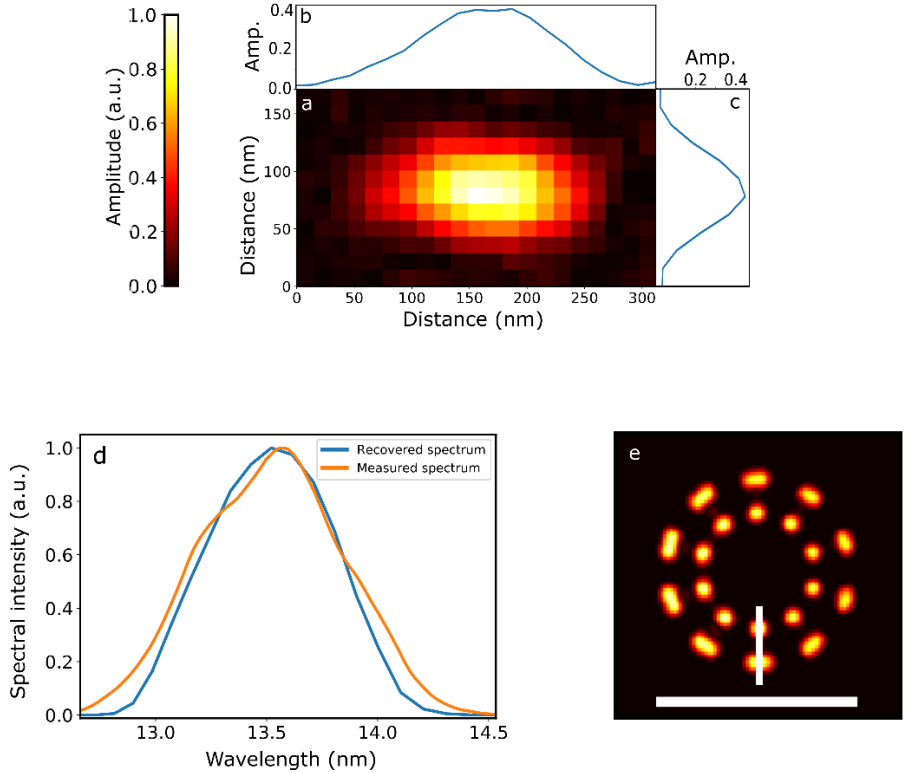

**Figure 4: Extraction of the spectrum and broadband phase retrieval. a** Shows the average of five cross-correlations of two reference apertures each. **b** and **c** show the lineouts of the averaged cross-correlation terms, which are averaged over the vertical and horizontal axis. **d** Spectrum obtained by deconvolving **b** with **c**, which is compared to the measured spectrum (orange). **e** Reconstructed object after applying a broadband, iterative phase retrieval algorithm on the measured diffraction pattern with Fig 3 e as the initial field and Fig 4 d as input spectrum. The scale bar in **e** corresponds to 1 µm.

In a second step, we refine the amplitude and phase image of the object by an iterative phase retrieval algorithm. Methods to reconstruct the phase for broadband diffraction patterns have already been developed, but require a precise measurement of the spectrum[25,33].

At this point we can benefit from the additional information encoded in the hologram again, namely, we can extract the spectrum from it. For this purpose, we investigated the cross-correlations of the reference apertures (see Fig. 2 **c**), which are smeared by the bandwidth in the same way as the sample is. An average of multiple smeared cross-correlation terms is shown in Fig. 4 **a**. The smearing can be described by a convolution of the reference aperture cross-correlation term with the spectrum used. Therefore, the spectrum can be recovered by a deconvolution procedure described in detail in the methods section. The recovered spectrum is shown in Fig. 4 **d** (blue) and agrees remarkably well with the measured spectrum (orange).

In the next step, the recovered spectrum and the result of the broadband FTH method are seeded into a broadband iterative phase retrieval algorithm. The result of this phase retrieval is shown in Fig. 4 **e** and shows nearly perfect image quality. A lineout along two separate features is shown in Fig. 5 and shows even higher resolution compared to the standard FTH method and the broadband FTH method. We estimate the improved resolution to be 34 nm using the 1/e criterion on the phase retrieval transfer function[34] (PRTF), which is very close to the Abbe-limit of 25 nm. Since the algorithm was started with a good initial estimate of the sample exit wave, the phase retrieval algorithm converges stably and quickly. Although the diffraction pattern was recorded with a high dynamic range and sufficient oversampling, an unseeded phase retrieval failed. Note that for the employed phase retrieval, all necessary input data including the spectrum are obtained from a single measured hologram. Thus, the presented holography-enhanced coherent imaging method can still be used in case of significant pulse to pulse fluctuations of the spectrum e.g. at unseeded free-electron-lasers or single-shot laser experiments[35].

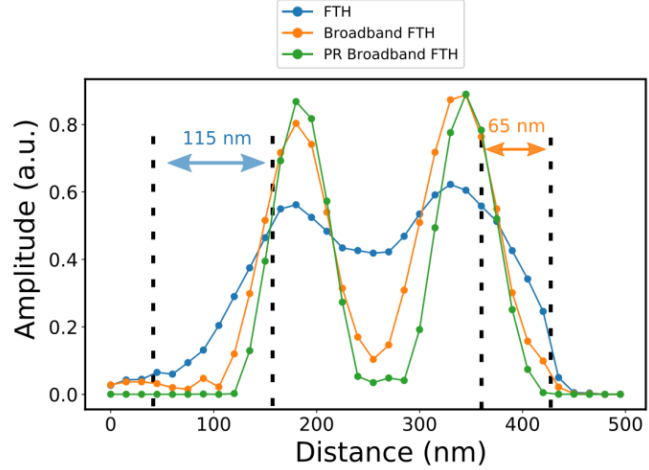

**Figure 5: Comparison of broadband FTH with standard FTH.** Lineouts along the white-line in Fig. 3 **d, e** and Fig. 4 **e**. The achieved resolution was estimated using a knife-edge on the standard FTH reconstruction and broadband-FTH reconstruction.

**Conclusion**

In summary, we demonstrated holography-enhanced coherent diffraction imaging with a broadband high harmonic source. A broadband FTH method, which relies solely on Fourier filters and is stable, robust, and fast was put into practice. Using an energy bandwidth of 5.5 eV, which corresponds to a relative energy bandwidth of 17, we demonstrated a spatial resolution of 65 nm, beyond the temporal coherence limit. Moreover, we have shown that the broadband spectrum can be recovered from the measured hologram as well, which allows seeding the spectrum and the broadband FTH result to a broadband iterative phase retrieval algorithm.

Combining these methods, for the first time, a nearly-diffraction limited spatial resolution of 34 nm has been demonstrated with a bandwidth supporting pulses as short as 380 as. This opens exciting possibilities for the observation of ultrafast transport and transfer processes of energy[28], charges[15], or spins[29], which are the basis of next-generation electronics, data storage, energy conversion, and energy storage devices.

The presented approach is particularly robust since a good starting point for the iterative algorithm as well as the required spectrum are extracted from a single measured hologram. Thus single-shot imaging gets feasible and shot-to-shot fluctuations can be tolerated. It can be applied at any broadband coherent light source including high harmonic beamlines, synchrotrons, and attosecond free-electron lasers.

**Methods**

*Sample fabrication:*

A $Si_3N_4$ membrane with a thickness of 50 nm was coated with 250 nm Cu, which results in an overall transmission of $4.5 \times 10^{-7}$ at 92 eV and acts as an absorber. The resolution test sample and reference apertures were structured by means of focused gallium ion beam milling on a FEI Helios NanoLab G3 UC. The resolution test sample and the corresponding reference apertures were written in parallel using an ion energy of 30 keV and current of 7.7 pA to achieve the high aspect ratio of the holes and bridges. After fabrication, the size of the reference apertures was measured using a transmission electron microscope yielding a diameter of 90 nm. The sample diameter was chosen to be 1 µm. The reference pinholes were placed at a radius of 2 µm equally spaced around the sample. Since the resolution imposed by the bandwidth depends on the maximal size of the sample (see eq. 1), the reference aperture should be placed as close as possible to the object. However, certain design rules have to be followed, which are explained in more detail in the supplementary material.

*Experiment and data processing*

A high-power fiber laser with a central wavelength of 1 µm was compressed to a pulse duration of 7 fs with a pulse energy of 400 µJ at a repetition rate of 76 kHz which results in an average power of 30 W. The IR-laser is focused in a gas jet with a diameter of 500 µm where argon with a backing pressure of 0.6 bar is applied. A broadband spectrum reaching up to 100 eV is generated with a flux of $7 \times 10^9$ phot/s/eV at 92 eV. A more detailed description of the XUV-source is given in[36]. To separate the high-power IR laser beam from the XUV-beam 4 grazing incident plates that reflect the XUV radiation, but transmit the IR beam, followed by two Zr filters with a thickness of 200 nm are used. Three broadband XUV-multilayer mirrors with a peak reflectivity each of 30 % at 92 eV are used to select a bandwidth of 5.5 eV (FWHM) and to focus the beam on the holography sample. The spectrum incident on the sample (Fig. 1 **b**) was calculated by multiplication of the measured spectrum of the HHG source, which was measured in an independent measurement using an XUV spectrometer, with the calibrated reflectivity of the three multilayer mirrors. The XUV-camera (Andor iKon-L) was placed 30 mm behind the sample which results in a NA of 0.42, but photons above the noise level were measured only up to an NA of 0.27, which results in a diffraction-limited resolution of 25 nm. The noise level was estimated by calculating the standard deviation of an area of the diffraction pattern where no signal was detected and equals for our measurement five times this measured standard deviation. During the measurements, the CCD was cooled down -50°C. The hologram in Fig. 1 **d**) was recorded with a total exposure time of 120 s using 2 by 2 on-chip binning. After the measurement, the hologram was corrected for the curvature on the Ewald's sphere, which is due to the high numerical aperture[37]. The resolution limit $\Delta r$, imposed by the bandwidth of the source was estimated using[25]

$$\Delta r = \lambda D/(2 l_c)$$

where D corresponds to the largest extend of the sample (2.5 µm for a single reference aperture) and $l_c$ to the lateral coherence length, which was approximated by the coherence length of a Gaussian-shaped spectrum ( $l_c = \sqrt{2 ln(2)}/\pi * \lambda^2/\Delta\lambda$ ) and results in a bandwidth limited resolution of 110 nm.

*Recovery of the spectrum*

In order to reconstruct the spectrum the cross-correlation of two reference apertures were investigated (Fig. 2 **a**). To increase the signal-to-noise ratio we averaged over several cross-correlations. For this purpose, the smeared cross-correlation terms were isolated, rotated by an angle of 36°, 72°, 108°, and 144° respectively and added up. The smearing along the horizontal axis can be described in a good approximation by a convolution of the monochromatic case with the spectrum. A detailed mathematical description can be found in the supplementary material. Hence, the spectrum can be recovered by a deconvolution, if the convolution kernel (i.e. the monochromatic cross-correlation of two reference apertures) is known. The 1D line profile of the smeared cross-correlation term, which is shown in Fig. 4 **b**, was extracted by calculating the mean value along the vertical axis. The next step is to estimate the convolution kernel. Since the reference aperture has a circular shape and the smearing along the vertical axis can be neglected, the vertical profile of the cross-correlation can be used as an approximation of the convolution kernel. To improve the signal-to-noise ratio, the convolution kernel, as it is shown in Fig 4 **c**, was approximated by the average along the horizontal axis. Finally, the spectrum was extracted by deconvolving the vertically averaged horizontal lineout (Fig. 4 **b**) by the horizontally averaged vertical lineout (Fig. 4 **c**). The deconvolution was calculated by using the Richardson-Lucy algorithm[38,39].

*Broadband, iterative phase retrieval algorithm*

The broadband-FTH result was seeded as the initial field to the broadband[25] version of the RAAR-algorithm[40] in combination with the shrink-wrap algorithm. The initial support was estimated by applying a threshold on the broadband-FTH reconstruction. Next, we ran the shrink-wrap algorithm for 5 iterations where we updated the support by applying a threshold on the convolution with a Gaussian kernel. We used a threshold of 0.2 for the inner part, where the object is located and a threshold of 0.05 at the area where the pinhole is located. The different limits were used because the exposure of the reference apertures was much weaker than the exposure of the sample area. After we obtained the support we ran the algorithm for 200 iterations using a beta parameter of 0.95 and updated the support every 20 iterations. The achieved resolution was evaluated with the PRTF (supplementary material), which results in a resolution of 34 nm.

**Data Availability**

The data that support the plots within this paper and other findings of this study are available from the corresponding author upon reasonable request.

**Acknowledgments**

This work was supported by the Federal State of Thuringia (2017 FGR 0076), the European Social Fund (ESF), the Thüringer Aufbaubank (TAB) for funding the junior research group HOROS (FKZ: 2017 FGR 0076), the European Research Council (ERC) under the European Union's Horizon 2020 research and innovation programme (grant agreement No [835306], SALT) and the Fraunhofer Cluster of Excellence Advanced Photon Sources. Further, this work was partially funded by the Center of Advanced Systems Understanding (CASUS) which is financed by Germany's Federal Ministry of Education and Research (BMBF) and by the Saxon Ministry for Science, Culture and Tourism (SMWK).

**Author Contributions**

W.E., S.W., C.L. and J.R. performed the imaging experiments. W.E., S.W. and J.R. analyzed the data. M.S. fabricated the sample and acquired the electron microscope image. S.Y. fabricated the broadband XUV mirrors. W.E., S.W. and H.M. implemented the broadband, iterative phase retrieval algorithm and M.B.,T.P., J.L. and J.R. initiated the project and designed parts of the experiment. All authors contributed to writing the manuscript.

**Competing Financial Interests**

The authors declare no competing financial interests.

# Supplementary information:

# Broadband holography-assisted coherent imaging – towards attosecond imaging at the nanoscale

Wilhelm Eschen [1,2], Sici Wang[1,2], Chang Liu[1,2], Michael Steinert[1], Sergiy Yulin[3], Heide Meissner[4], Michael Bussmann[4,5], Thomas Pertsch[1,3], Jens Limpert[1,2,3] & Jan Rothhardt [1,2,3]

**Mathematical description of Fourier transform holography with broadband illumination:**

Let us consider the measurement of the diffraction pattern $I_m$ with the exposure time T on a spatial detector.

$$I_m(x',y') \propto \int_0^T |\Phi(t,x',y')|^2 dt$$

Here $\Phi$ is the scalar electric field measured by the detector. Since the exposure time in conventional experiments is much longer than the period of oscillation of the electromagnetic field, the theorem of Parseval can be applied, so that follows:

$$I_m \propto \int_0^\infty |\Phi(\omega)|^2 d\omega$$

Assuming the far-field approximation, $\Phi$ can be written in FTH by the Fourier transform of the object $O(x,y)$ and the spatially separated reference aperture $R(x,y)$.

$$\Phi(\omega,x',y') \propto \frac{\omega}{iz}\sqrt{\alpha(\omega)}\mathcal{F}[O(x,y)+R(x,y)]\left(q_x=\frac{x'\omega}{zc}, q_y=\frac{y'\omega}{zc}\right)$$

Here $\mathcal{F}$ represents the Fourier transform, $(q_x,q_y)$ the spatial frequencies and $(x',y')$ the coordinates in the detector plane. $\alpha(\omega)$ corresponds to the spectral weights of the spectrum used. Using this equation for the diffraction pattern the broadband far-field intensity $I_m$ can be written as:

$$I_m(x',y') \propto \int_0^\infty \frac{\omega^2}{z^2}\alpha(\omega)\left|\mathcal{F}[O(x,y)+R(x,y)]\left(\frac{x'\omega}{zc},\frac{y'\omega}{zc}\right)\right|^2 d\omega$$

In the next step, we consider only the cross-correlation term of the object with the reference aperture. Thus the following expressions results, where $\otimes$ represents the convolution operator.

$$I_m(x',y') \propto \int_0^\infty \frac{\omega^2}{z^2} \alpha(\omega) \mathcal{F}[O(x,y) \otimes R^*(x,y)]\left(\frac{x'\omega}{zc}, \frac{y'\omega}{zc}\right) d\omega$$

In the next step, the inverse Fourier transform is applied to the diffraction pattern. Considering the sampling $\Delta p$ of the hologram by the detector and the scaling property of Fourier transforms, the following expression appears.

$$\mathcal{F}^{-1}[I_m](n_x, n_y) \propto \int_0^\infty \alpha(\omega) O(n_x \Delta r, n_y \Delta r) \otimes R(n_x \Delta r, n_y \Delta r) \, d\omega \qquad (1)$$

Where $n_x$ and $n_y$ correspond to the pixel numbers and the realspace pixel size $\Delta r$ is determined by

$$\Delta r = \frac{2\pi z c}{N \Delta p \omega}$$

Where N is the number of pixels of the detector. Thus, the broadband reconstruction yields a superposition of many monochromatic reconstructions scaled by the wavelength (Fig 7 b), since the pixel size is scaled by $\lambda$. The broadband FTH method corrects this achromatic scaling to a large extent.

**Broadband Fourier transform holography method:**

Above it has been shown that a broadband spectrum leads to a superposition of many reconstructions scaled by wavelength. To illustrate this behavior, the sample used in our experiment was simulated (Fig. 6 a) assuming a monochromatic illumination and an illumination consisting of two discrete wavelengths located at 12.3 nm and 15.0 nm. In the monochromatic case (Fig. 6 b), the object is resolved and only blurred due to the size of the reference aperture. For the dual-wavelength case (Fig. 6 c), the reconstruction is distorted. Two shifted, partially overlapping objects are visible, where each reconstructed sample corresponds to a single wavelength. This behavior explains the anisotropic resolution since the object is shifted along the "sample-pinhole" axis, which has been exploited in earlier experiments where a specialized sample design was used to separate the contributions of different discrete wavelengths and achieve wavelength multiplexing[1].

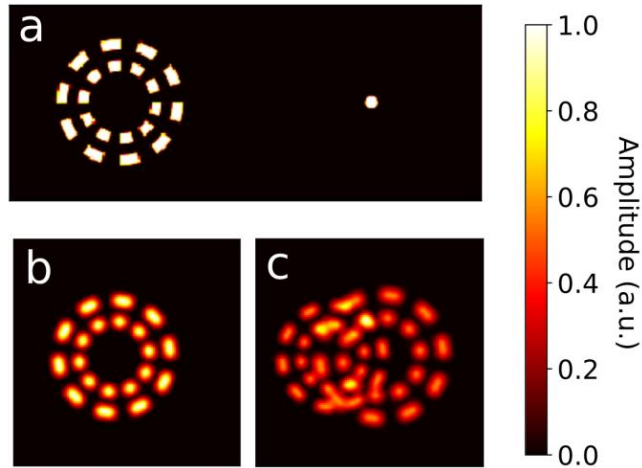

**Figure 6: Effect of broadband radiation on FTH**. **a** shows the simulated sample. **b** shows a monochromatic reconstruction and **c** shows the reconstruction for an illumination consisting of two discrete wavelengths.

**Sample design:**

The bandwidth limited spatial resolution in FTH depends linearly on the distance between the sample and the corresponding reference aperture[2]. Unfortunately, the reference aperture cannot be placed too close to the sample, since geometrical constraints have to be considered. At the center of the reconstruction the autocorrelation of the sample is located, which has a radius of two times the sample radius (see Fig. 7 a). Since the cross-correlation term (CC), which contains the image of the sample should not overlap with the auto-correlation term (AC), the reference needs to be placed at a distance of at least 3r from the sample center, where r is the radius of the sample. If a broad-band source is assumed, the situation gets a bit more difficult. Since the reconstruction is a superposition of multiple reconstructions with different wavelengths (see Fig. 7 b for illustration), the minimal distance increases and needs to fulfill the following condition[3]:

$$d \geq r\left(1 + \frac{2\lambda_{max}}{\lambda_{min}}\right) \quad (2)$$

Where $\lambda_{min}$ and $\lambda_{max}$ are the smallest and largest wavelength of the spectrum.

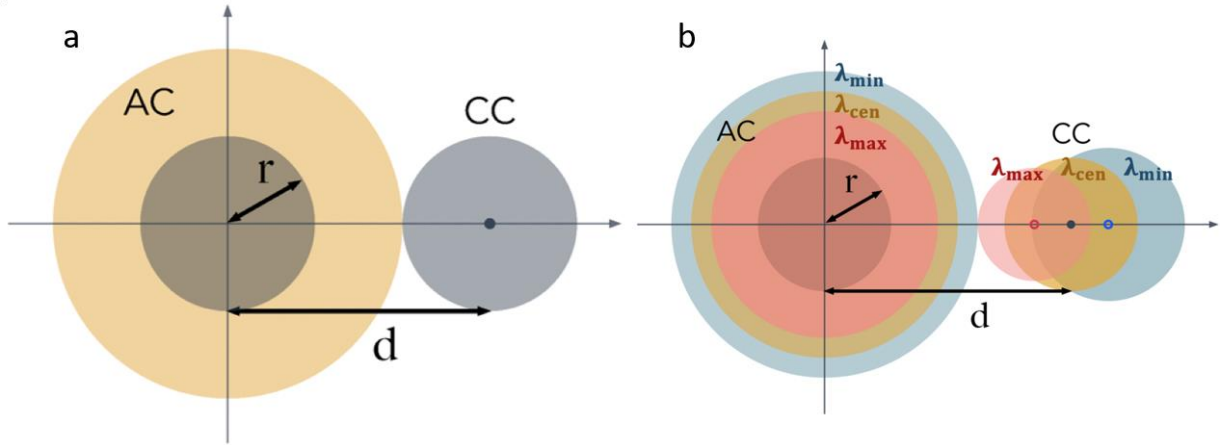

**Figure 7: Design constraints for FTH samples a**, Shows the monochromatic case. The autocorrelation (AC) is yellow and has the double size of the original sample. The cross-correlation term (CC) represents the reconstruction of the object. The broadband case is shown in **b**. Due to the broad bandwidth, the design rules have to be adapted.

**The number of reference apertures:**

Since the broadband-FTH method relies on selecting high-contrast Fourier components from multiple CC terms, adding as many reference apertures as possible is desirable. Unfortunately, not more than 5 pinholes can be added without increasing the sample-pinhole distance which is illustrated in Fig. 8. The reason for this is that otherwise, the cross-correlation terms will overlap. Overlapping cross-correlation terms can be avoided for more than 5 pinholes by using a larger pinhole-sample distance d. However, this will further increase the smearing effect on each CC-team due to the broad bandwidth. Please note that only half of the circle is filled with reference pinholes since each CC term will also have a complex conjugate on the opposite side of the sample.

Moreover, a larger distance also requires a larger beam and therefore more photons are wasted in the absorbing mask. Therefore we used 5 reference apertures in our experiment.

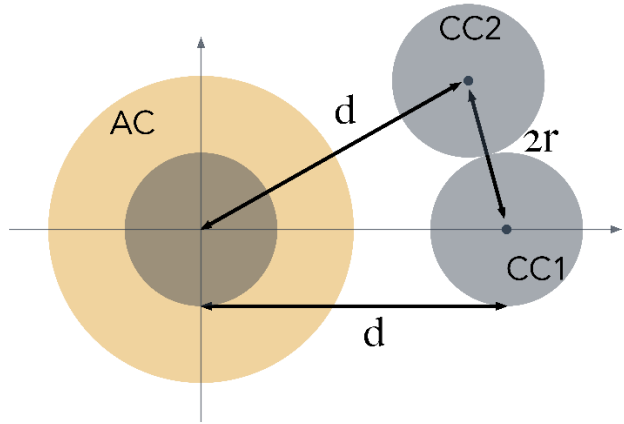

**Figure 8:** This figure illustrates that only 5 reference apertures can be used without increasing the pinhole-sample distance d.

**Limits of broadband-FTH:**

Here we answer the question to what extent broadband-FTH will improve the resolution compared to standard FTH. Above it was shown, that the reference aperture has to be placed at least $3 \cdot r$ from the center of the sample (for monochromatic radiation and infinite small reference aperture), where r is the radius of the sample. However, in our experiment, a broad bandwidth source was used and the reference aperture itself had a diameter of 90 nm. Besides generally, in FTH a good separation of the CC term from the AC term is important since the AC term is usually much brighter than the CC term. Thus $4 \cdot r$ can be considered as a reasonable sample-pinhole distance.

For the broadband case the smearing of the reconstruction is due to the superposition of many monochromatic reconstructions stretched by the wavelength (see Fig. 7 b for illustration). Our method removes the stretching due to the large pinhole-sample distance. The stretching, that is due to the extent of the sample itself cannot be removed. Hence, the broadband FTH method can improve the resolution for the given geometry by a factor of 5. These findings are proven by a numerical simulation. A Siemens star with a diameter of 2 µm, surrounded by five reference apertures at a radius of 4 µm, is used as the sample. The size of the reference apertures was chosen to be 80 nm. Furthermore, a Gaussian-shaped spectrum at a central photon energy of 92 eV with a relative bandwidth of dE/E=1/10 (FWHM) was assumed. The result of the standard-FTH method is shown in Fig. 9 c and the reconstruction of the broadband-FTH method is shown in Fig. 9 d. From the Siemensstar sample, the resolution can be estimated directly from the reconstruction by the smallest radius where the spokes are still resolved. As the resolution criterion, the Rayleigh criterion was chosen, which means a structure was considered as resolved if a contrast of at least 0.264 is visible. Our analysis shows that with the standard-FTH method all spokes are resolved at a radius of 837 nm (see Fig. 9 e), whereas with broadband-FTH all spokes are already resolved at a radius of 176 nm (see Fig. 9 f). Since the spacing of the spokes is linear with the radius, our simulation shows an improvement of the resolution by a factor of 4.8, which is in good agreement with our geometrical considerations.

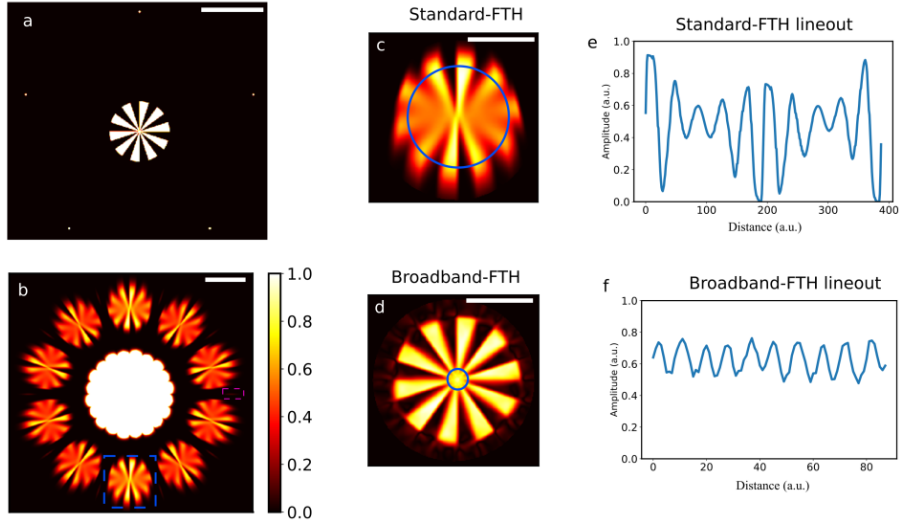

**Figure 9: Numerical simulation to determine the limit of broadband FTH. a**, Shows the sample (Siemens star) used in our simulation. The scale bar corresponds to 2 µm. **b**, Fourier transform of the simulated broadband hologram. The blue box highlights a single cross-correlation term of the Siemens star with a reference aperture. The purple box shows the corss-correlation of two reference apertures. **c,** The reconstruction of the standard FTH method shows poor resolution. **d,** Reconstruction of the broadband-FTH method. **e** Circular lineout indicated in **c** by a blue line. **f** Circular lineout indicated in **d** by a blue line.

**Recovery of the spectrum:**

The reconstruction of the hologram allows to recover the spectrum as well. The basic idea is that the spectrum can be obtained from the smeared cross-correlation term of two reference apertures ($R_{cc}$), that are separated by a distance $x_0$.

$$R_{cc}(x,y) = R_1(x - x_0, y) \otimes R_2(x,y)$$

Using the cross-correlation term of two reference apertures in equation 1 instead of the cross-correlation with the object, the following equation is obtained:

$$\mathcal{F}^{-1}[I_m](n_x, n_y) \propto \int_0^\infty \alpha(\omega) R_{cc}(n_x \Delta r - x_0, n_y \Delta r) \, d\omega$$

Assuming that the diameter of the reference aperture is small compared to the distance $x_0$ and that the bandwidth is not too large the following expression can be derived by using a Taylor expansion.

$$\mathcal{F}^{-1}[I_m](n_x, n_y) \propto \int_0^\infty \alpha(\omega) R_{cc}\left(\Delta r_c n_x - x_0 * \frac{\omega}{\omega_c}, n_y \Delta r_c\right) d\omega$$

Where $\omega_c$ is the central frequency of the spectrum and $\Delta r_c$ is the pixel size corresponding to this frequency. The right side of the resulting equation represents a convolution. Thus the smeared cross-correlation term of two reference apertures can be represented in good approximation by a convolution with the spectrum. This means that if the shape of the reference aperture is known, the spectral shape can be recovered by a deconvolution.

Here this method is demonstrated for the simulation shown above. The cross-correlation from the simulation in Fig. 9 (purple box) was extracted and plotted in Fig. 10 a. The cross-correlation of the reference aperture used here is smeared only along the horizontal direction, which is marked by a white arrow. The spectrum can be recovered by a deconvolution if the convolution kernel (i.e. the cross-correlation of the pinholes) is known. The 1D line profile of the smeared cross-correlation term was estimated by calculating the mean value along the vertical axis (Fig. 10 b). Since the reference aperture has a circular shape, we can use the vertical profile of the cross-correlation as the convolution kernel. Hence, the convolution kernel was approximated by calculating the mean value along the horizontal axis, which is displayed in Fig. 10 c. The spectrum was extracted by deconvolving the horizontal average (Fig. 10 b) by the vertical average (Fig. 10 c). The initial spectrum, which was used for the simulation, and the reconstructed spectrum are shown in Fig. 10 d and agree very well.

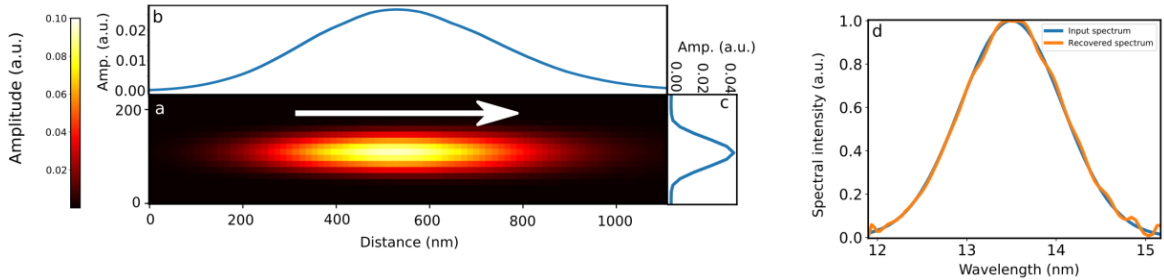

**Figure 10: Reconstruction of the spectrum from the smeared cross-correlation term of two reference apertures a**, A single cross-correlation of two reference apertures, which was extracted from Fig. 9 (purple box) and is smeared due to the large bandwidth. **b**, Average of **a** along the vertical axis. **c,** Average of **a** along the horizontal axis. **d**, Input spectrum of the simulation shown in Fig. 9 and reconstructed spectrum using the method described in this section.

## Resolution of the phase retrieval results:

The resolution for the standard FTH method and the broadband FTH method was estimated by the steepness of an edge. Since waveguiding effects that occur during propagation inside the sample lead to smooth edges and hence can lead to an underestimated resolution[5,6], the PRTF[4] was used to determine the resolution achieved by the phase retrieval algorithm. For this purpose an PRTF adopted the to the broadband case was used:

$$\text{PRTF}(q) = \frac{\sqrt{\sum_\lambda a(\lambda)|\psi(q,\lambda)|^2}}{\sqrt{I(q)}}$$

Where $a(\lambda)$ corresponds to the spectral weighting factor and $\psi(q,\lambda)$ to the exit wave propagated to the far-field, hence the numerator corresponds to the reconstructed diffraction pattern and the denominator to the square root of the measured diffraction pattern. We applied the 1/e criterion for the estimation of the resolution which corresponds to the maximal spatial frequency of 14.7 µm$^{-1}$ (see Fig. 11) and hence to a resolution of 34 nm.

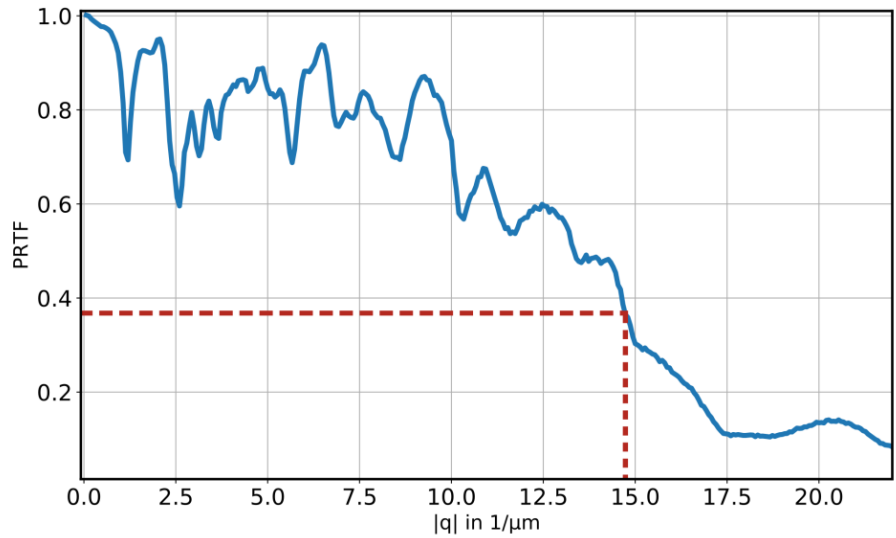

**Figure 11: Phase retrieval transfer function (PRTF).** The PRTF shows that spatial frequencies up to 14.7 µm$^{-1}$ were successfully phased. Hence, we achieved a resolution of 34 nm using a broadband phase retrieval algorithm.